\documentclass[journal]{IEEEtran}
\usepackage[T1]{fontenc}
\usepackage[utf8]{inputenc}
\usepackage{cite}
\usepackage{amsmath}
\usepackage{array}
\usepackage{booktabs}
\usepackage{graphicx}
\usepackage{tikz}
\usepackage{xcolor}
\usepackage{url}
\usepackage{balance}
\usepackage{hyperref}
\usepackage{setspace}

\usetikzlibrary{arrows.meta,positioning,fit,calc,shapes.geometric}

\title{In-Orbit Computing for Remote Sensing: Task-Oriented Processing and Transmission}

\author{Yi Zhang, Zesong Fei,~\IEEEmembership{Senior~Member,~IEEE}, Jingxuan Huang,~\IEEEmembership{Member,~IEEE}, Hua Dang, Jihao Luo, \\Wenyu Liang, Haoran Zhang, Xinyi Wang,~\IEEEmembership{Member,~IEEE}
	
	\thanks{Yi Zhang, Zesong Fei, Jingxuan Huang, Jihao Luo, Wenyu Liang, Haoran Zhang and Xinyi Wang are with the School of Information and Electronics, Beijing Institute of Technology, Beijing 100081, China (e-mail: zhangyizy0776@163.com, feizesong@bit.edu.cn, jxhbit@gmail.com, jihaoluo$\_$bit@icloud.com, 13360736036@163.com, 7220210164@bit.edu.cn, wangxinyi@bit.edu.cn).}
	\thanks{Hua Dang is with the School of Integrated Circuits and Electronics, Beijing Institute of Technology, Beijing 100081, China (e-mail: danghua@bit.edu.cn).}
\vspace{-15pt}
}

\begin{document}
\maketitle

\begin{abstract}
The growing volume and timeliness requirements of remote sensing data are exposing the limitations of the conventional “transmit-then-process” architecture, particularly under intermittent satellite-ground connectivity, constrained downlink bandwidth, and limited onboard computing resources. This article presents a semantic-aware space-ground collaborative remote sensing (SSGC-RS) framework, in which remote sensing data are processed to different semantic depths according to task requirements and available communication-computing resources. Three operating modes are considered, ranging from fully onboard task inference to semantic representation delivery and space-ground split inference. A land-cover recognition case study illustrates how these modes address different communication and computing bottlenecks. Finally, emerging challenges and future directions for intelligent remote sensing systems are discussed.

\end{abstract}


\section{Introduction}

The rapid development of satellite constellations and persistent Earth observation capabilities is generating unprecedented volumes of remote sensing (RS) data for applications such as flood monitoring, disaster response, precision agriculture, ecological assessment, and urban management \cite{xin2026orbit}. At the same time, many of these applications are becoming increasingly time-sensitive, requiring useful information to be delivered shortly after observation \cite{barretta2026toward, jiang2025high}. 
However, conventional RS systems still largely follow a ``transmit-then-process'' paradigm, where satellites acquire and compress raw observations, while computationally intensive tasks such as segmentation, detection, and change analysis are performed after the data reach the ground. Such an architecture is effective for offline analysis and data archiving, but its limitations become pronounced when downlink bandwidth is scarce, satellite-ground contact windows are short, and mission deadlines are stringent. The resulting chain of data compression, transmission, reconstruction, and ground processing may introduce considerable latency, while substantial communication resources are consumed by information that may ultimately be irrelevant to the target task~\cite{sang2026onboard}. 

The continuing improvement of onboard computational capabilities is beginning to change this processing paradigm. Recent in-orbit artificial intelligence demonstrations have shown that satellites can perform increasingly sophisticated data analysis before downlink~\cite{del2024mitigating}, while missions such as $\Phi$sat-2 illustrate the potential of onboard intelligence processing for timely Earth observation and environmental monitoring~\cite{razzano2024ai}. Meanwhile, semantic communication and task-oriented inference have motivated a transition from delivering complete observations toward transmitting information according to its relevance to downstream tasks~\cite{jiang2025high}. These advances make it practical to execute part of the remote sensing processing chain before downlink, rather than treating the satellite only as a sensing and forwarding platform.

However, pushing intelligence onboard does not imply that all RS tasks should be completed entirely in space. Appropriate processing depth depends on the mission bottleneck. When the downlink opportunity is extremely limited or a rapid task response is required, it may be preferable to complete task-specific inference onboard and transmit only compact results. When richer information must remain available to ground centers, transmitting compressed semantic representations may provide a better balance between information richness and communication cost. Conversely, when complex models exceed onboard computing or energy budgets, part of the inference can be retained on the satellite while the remaining computation is completed on the ground. Therefore, the central point is not simply whether intelligence should be moved onboard, but rather how much task-relevant information should be extracted in space.

\begin{figure*}[t]
\centerline{\includegraphics[width=2\columnwidth]{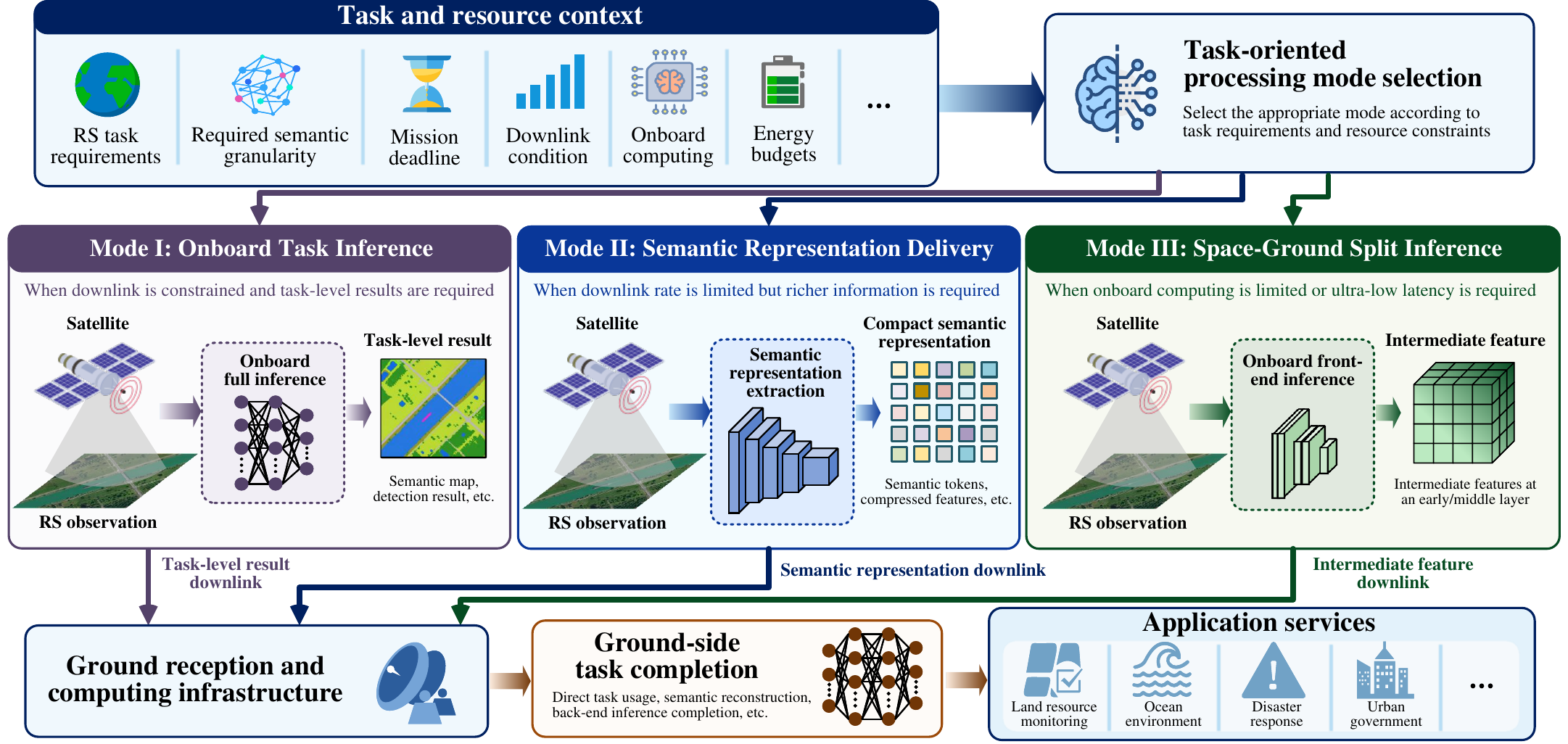}}
\caption{Task- and resource-aware SSGC-RS framework with three representative operating modes.}
\label{fig1}
\end{figure*}

From this perspective, semantic extraction provides a key interface between in-orbit computing and space-ground communication. Instead of treating all acquired pixels equally, semantic extraction transforms raw observations into representations at different levels of task relevance and information richness, ranging from task-level results to compact semantic representations and intermediate model features. The appropriate representation depends jointly on mission requirements and available communication-computing resources. This motivates a semantic-aware space-ground collaborative remote sensing (SSGC-RS) framework, in which task requirements, latency constraints, downlink conditions, visibility windows, onboard computing and energy resources, and model characteristics jointly determine the processing mode adopted for each mission.


This article develops the SSGC-RS framework from a task- and resource-aware processing perspective. The article details the architectural design of the framework and shows how task and resource context drive the selection of an operating mode. Section III then examines three representative operating modes for task-oriented space-ground inference: onboard task inference, semantic representation delivery, and space-ground split inference. We use a land-cover recognition case study to illustrate how the three modes behave under different task requirements and resource constraints, and then discuss the remaining challenges in adaptive mode selection, trustworthy semantic extraction, and constellation-scale coordination.

\section{System Architecture}

\subsection{Framework Overview}
Fig.~\ref{fig1} illustrates the proposed SSGC-RS framework, which treats onboard processing depth as a decision variable rather than a fixed system configuration. Given a remote sensing task, the system determines how much processing is performed in orbit, what representation is transmitted over the space-ground link, and what processing remains on the ground.

The operating modes are determined by two types of information. Task context specifies what has to be ultimately delivered, e.g., a rapid decision, a reusable scene representation, or a complete inference result, while resource context determines how much processing can be executed onboard and how much information can be carried during the available contact window. Their combination determines both the semantic depth of onboard processing and the form of the downlink payload.

This design shifts space-ground remote sensing from a task-agnostic data pipeline toward adaptive task-oriented inference, where the amount and form of information transmitted to the ground are jointly determined by task value and resource availability.

\subsection{Task- and Resource-Aware Operating Modes}
Depending on the different combinations of task requirements and available communication and computing resources, the SSGC-RS framework selects among three representative operating modes, as illustrated in Fig.~\ref{fig1}.

\textit{1) Onboard task inference} suits missions in which the downlink or the visibility window is the binding constraint while the onboard platform can host the required model within its computing and energy budget. The satellite completes the whole inference and downlinks task-level results such as a semantic map, a change map, or a list of detected objects. The transmission load is far smaller than the observation, but the ground loses access to visual information that was discarded by the in-orbit processing.

\textit{2) Semantic representation delivery} suits missions in which the downlink rate is limited but the ground still needs richer information than a task-level result, for instance to support fusion, fine-grained analysis, or archiving. The satellite extracts and compresses a compact semantic representation of the scene and downlinks that representation, and the ground decodes it or reconstructs the scene from it. Its usefulness therefore depends on whether the compact representation preserves the information required for subsequent ground processing at the available downlink rate.

\begin{table*}[htbp]
	\centering
	\caption{Comparison of Task- and Resource-Aware Operating Modes}
	\label{tab:use_cases}
	\fontsize{6.6}{8.1}\selectfont
	\setlength{\tabcolsep}{5pt}
	\setlength{\parskip}{0pt}
	\setlength{\parindent}{0pt}
	\renewcommand{\arraystretch}{1.16}
	\begin{tabular}{|>{\raggedright\arraybackslash}p{84pt}|>{\raggedright\arraybackslash}p{104pt}|>{\raggedright\arraybackslash}p{94pt}|>{\raggedright\arraybackslash}p{92pt}|>{\raggedright\arraybackslash}p{89pt}|}
		\hline
		\multicolumn{1}{|>{\centering\arraybackslash} p{84pt}|}{\textbf{Operating Mode}} &
		\multicolumn{1}{>{\centering\arraybackslash} p{104pt}|}{\textbf{Preferred Conditions}} &
		\multicolumn{1}{>{\centering\arraybackslash} p{94pt}|}{\textbf{Key Design Considerations}} &
		\multicolumn{1}{>{\centering\arraybackslash} p{92pt}|}{\textbf{Main Advantage}} &
		\multicolumn{1}{>{\centering\arraybackslash} p{89pt}|}{\textbf{Main Limitation}} \\ \hline

		\textbf{Onboard Task Inference} &
		Limited downlink resources; time-sensitive and task-specific output &
		Onboard model complexity and task-result reliability &
		Low transmission load and direct delivery of task results &
		Limited flexibility for subsequent ground-side analysis \\ \hline

		\textbf{Semantic Representation Delivery} &
		Limited downlink resources; richer scene information required &
		Representation compactness and preservation of task-relevant information &
		Reduced transmission rate while retaining reusable scene information &
		Information loss due to compression and semantic selection \\ \hline

		\textbf{Space-Ground Collaborative Inference} &
		Limited onboard computing capacity; sufficient downlink capacity for intermediate feature transmission &
		Model partitioning and communication-computing resource allocation &
		Reduced task latency through distributed processing &
		Performance depends on feature size and available resources \\ \hline

	\end{tabular}
\end{table*}

\textit{3) Space-ground collaborative inference} suits missions in which the onboard computing or energy budget cannot support the whole model, or in which a complex model must still meet a tight latency deadline. The satellite executes the front-end part of the network and downlinks intermediate features, and the ground station completes the remaining inference. Computing and communication load are shared between the two segments, and the split point jointly determines onboard computation, downlink volume, and the remaining ground workload.

To conclude, onboard task inference and semantic representation delivery both address constrained downlink resources but differ in the information required by ground-side applications. The former is suitable when task-level outputs are sufficient, whereas the latter retains richer scene information for subsequent processing. Space-ground collaborative inference instead addresses limited onboard computing capacity by distributing model execution between satellite and ground nodes, provided that intermediate features can be transmitted efficiently. 

Regardless of the mode, downlink content enters the same ground reception and computing infrastructure, while the required ground operation differs: direct use of a task result, decoding or reconstruction of a semantic representation, or completion of back-end inference. Table~\ref{tab:use_cases} summarizes the preferred conditions, key design considerations, main advantage, and main limitation of the three modes.

\section{Three Operating Modes for Task-Oriented Space-Ground Inference}
The modes introduced in Section II differ in how much of the task is completed on each side.

\subsection{Onboard Task Inference: From Raw Observation to Task-Level Results}
Onboard task inference places the whole inference chain on the satellite and sends only its outcome, which is attractive when the task result is much smaller than the source observation and the onboard platform can execute the required model within its latency and energy budget. In this case, transmitting the raw scene wastes scarce contact windows on information that is not needed by the current task.

The downlink content depends on the task: a class map for semantic segmentation, object locations and labels for detection, or a change mask for change analysis. Since the underlying pixels may no longer be available on the ground, the task processing result should ideally be accompanied by a confidence measure that identifies regions requiring further inspection.

\begin{figure}[t]
\centerline{\includegraphics[width=0.9\columnwidth]{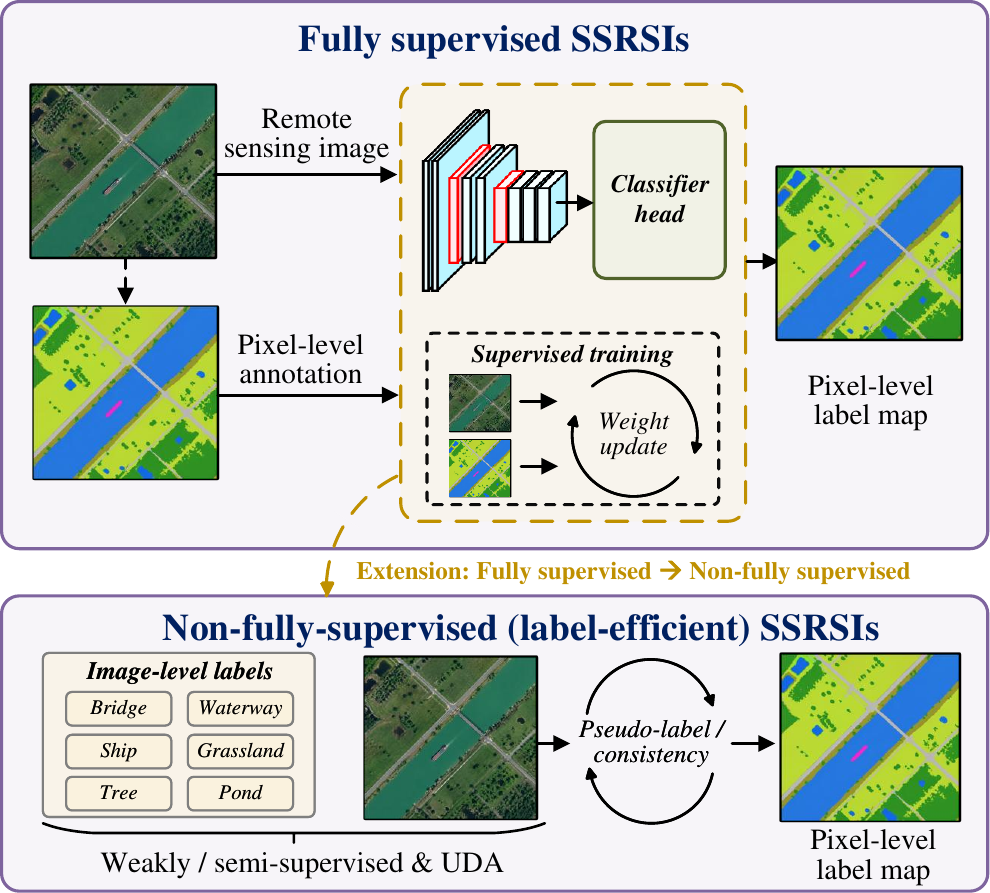}}
\caption{Framework comparison between fully supervised and non-fully-supervised SSRSIs.}
\label{fig2}
\end{figure}

Semantic segmentation of remote sensing images (SSRSIs) provides a representative example of onboard task inference because its output, i.e., a pixel-level class map, is substantially more compact than the source image while remaining useful for land-cover applications. Among the various models for SSRSIs, fully supervised models remain the dominant choice because they generally provide the most reliable pixel-level predictions when sufficient annotations are available. Their performance benefits from a series of advances in segmentation architectures, including fully convolutional networks for end-to-end pixel prediction, encoder-decoder structures with skip connections for spatial-detail recovery, atrous convolution and pyramid pooling for multi-scale context aggregation, and more recently transformer-based models for long-range dependency modeling \cite{huang2023deep}. These capabilities are particularly important for remote sensing, which often exhibits large intra-class variation, high inter-class similarity, complex backgrounds, and substantial scale changes across objects.

\begin{table*}[htbp]
	\centering
	\caption{Comparison of Pixel-Based and Token-Based Transmission Strategies }
	\label{token-comm-comparison}
	\fontsize{6.6}{8.1}\selectfont
	\setlength{\tabcolsep}{5pt}
	\setlength{\parskip}{0pt}
	\setlength{\parindent}{0pt}
	\renewcommand{\arraystretch}{1.16}
	\begin{tabular}{|>{\raggedright\arraybackslash}p{90pt}|>{\raggedright\arraybackslash}p{96pt}|>{\raggedright\arraybackslash}p{90pt}|>{\raggedright\arraybackslash}p{92pt}|>{\raggedright\arraybackslash}p{95pt}|}
		\hline
		\multicolumn{1}{|>{\centering\arraybackslash} p{90pt}|}{\textbf{Communication Scheme}} &
		\multicolumn{1}{>{\centering\arraybackslash} p{96pt}|}{\textbf{Representation}} &
		\multicolumn{1}{>{\centering\arraybackslash} p{90pt}|}{\textbf{Rate Adaptation}} &
		\multicolumn{1}{>{\centering\arraybackslash} p{92pt}|}{\textbf{Awareness}} &
		\multicolumn{1}{>{\centering\arraybackslash} p{95pt}|}{\textbf{Main Limitation}} \\ \hline

		\textbf{Conventional Image Transmission} &
		Compressed pixel-domain bitstream &
		Quantization and coding parameters &
		Not explicitly task-aware &
		Loss of task-relevant details at low bitrates \\ \hline

		\textbf{Unified Token Transmission} &
		Discrete token indices from a learned codebook &
		Token resolution and coding rate &
		Uniform token treatment &
		No explicit prioritization of important tokens \\ \hline

		\textbf{Semantic-Aware Token Transmission} &
		Importance-weighted semantic tokens &
		Importance-guided token allocation &
		Region- or task-dependent prioritization &
		Sensitive to inaccurate semantic importance estimation \\ \hline

	\end{tabular}
	
\end{table*}

The main limitation of fully supervised SSRSIs is its dependence on large amounts of pixel-level annotation, which is costly to obtain and difficult to maintain as new geographic regions, sensors, and observation conditions are introduced. Label-efficient approaches therefore provide an attractive alternative for large-scale and continuously evolving remote sensing applications. Weakly supervised methods reduce annotation efforts by learning from coarse labels such as image-level tags, scribbles, or bounding boxes, often with the aid of class activation maps. Semi-supervised methods combine a limited amount of labeled data with abundant unlabeled imagery through pseudo-labeling or consistency regularization. Unsupervised domain adaptation further addresses distribution shifts by transferring knowledge from a labeled source domain to an unlabeled target domain, thereby improving robustness across different sensors, regions, or seasons \cite{huang2023deep}. Compared with fully supervised training, these label-efficient paradigms reduce the annotation and update cost of onboard models, which is important for remote sensing systems that must operate across changing environments and mission requirements. Their main drawback is that the resulting segmentation accuracy and reliability generally remain below those of fully supervised models. Fig.~\ref{fig2} summarizes the difference between the two families.

The main limitation of this mode is irreversibility. If the onboard model misses an object or discards a region, the ground cannot recover that information without an additional transmission.

\subsection{Semantic Representation Delivery: From Pixel Transmission to Compact Semantics}
A task-level result is insufficient when the ground must support subsequent fusion, re-analysis, or archival use. In such cases, the downlink should retain more scene information without reverting to full-resolution pixel transmission. Conventional image codecs are primarily designed to preserve visual fidelity rather than downstream task performance. Consequently, their bit allocation may not adequately reflect the importance of different regions or objects to the remote sensing task. 

To reduce the downlink payload while retaining information for subsequent ground-side processing, remote sensing observations can be converted into compact representations before transmission. These representations range from region-level descriptions, such as object masks and proposals, to learned features that preserve richer scene information. Among them, discrete token communication first maps the image to a latent feature grid, where each feature vector is quantized to an entry in a learned codebook, and the resulting token indices are then transmitted as a compact representation of the original observation. The source rate can be adjusted through token resolution, coding, or selective token retention, and these representations offer a practical interface to digital communication systems. 

Tokens can be transmitted uniformly or selected according to their semantic importance. Unified token transmission treats all tokens equally and relies on the codebook and the decoder for compression and reconstruction; a representative realization is the vector-quantized generative adversarial network (VQ-GAN), which combines vector quantization with adversarial and perceptual training and a learned prior over token sequences \cite{esser2021taming, qiao2025token}. Semantic-aware token transmission instead uses the importance of different image regions to guide token selection. For example, a semantic mask can be combined with vector quantization to retain important tokens while masking less relevant ones, reducing the transmission rate while preserving task-relevant information \cite{pezone2025sq}. Beyond source representation, semantic importance can also guide transmission protection: critical tokens may receive stronger channel coding, more robust modulation, or higher transmission priority when supported by the communication system \cite{zhang2026tokencom}. Table~\ref{token-comm-comparison} compares conventional image transmission with unified and semantic-aware token transmission. 

The effectiveness of semantic-aware token transmission depends on the accuracy of importance estimation. If important regions are overlooked, their associated tokens may be discarded, resulting in information loss that cannot be recovered during reconstruction. Such errors may remain hidden under conventional rate-distortion metrics while substantially degrading downstream task performance. 

\begin{figure}[t]
\centerline{\includegraphics[width=0.9\columnwidth]{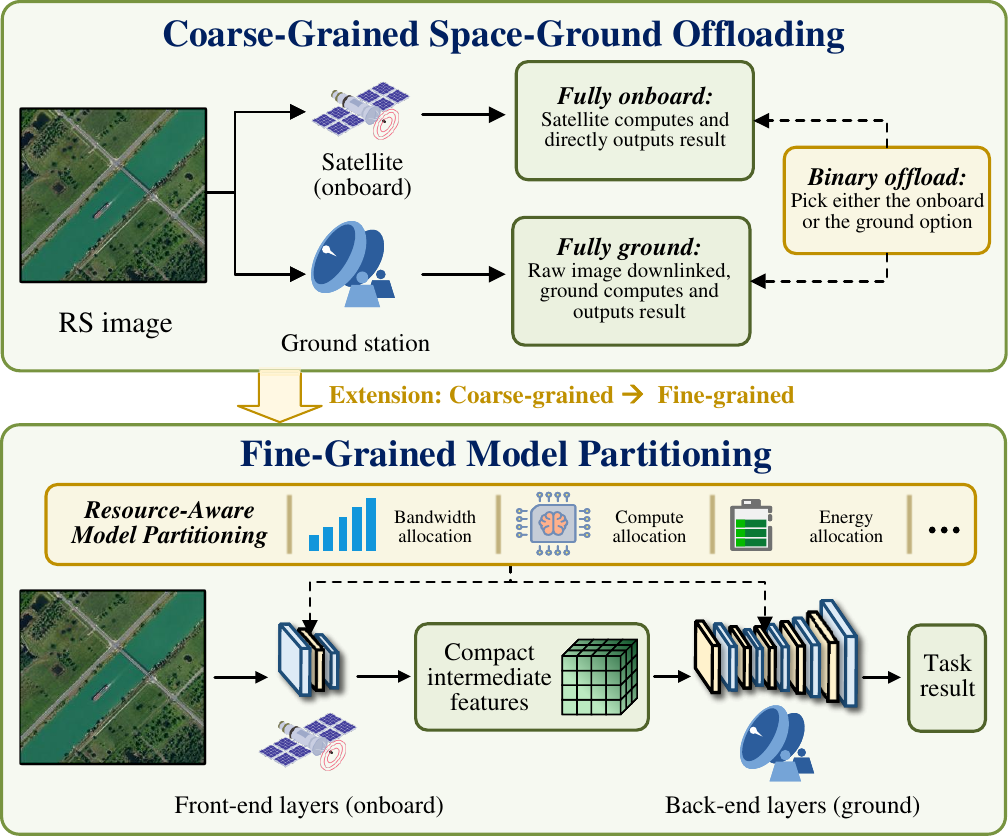}}
\caption{Illustration of coarse-grained space-ground offloading and fine-grained, resource-aware model partitioning for space-ground split inference.}
\label{fig3}
\end{figure}

\subsection{Space-Ground Collaborative Inference: Balancing Communication and Computing}
Space-ground collaborative inference applies when the onboard platform cannot carry the whole model within its computing or energy budget, or when a complex model must still meet a tight deadline. The network is divided, the satellite executes the early part, and the ground completes the rest.

Fig.~\ref{fig3} contrasts coarse-grained offloading with fine-grained model partitioning. The former treats the task as a whole and makes an all-or-nothing decision. Fully ground execution downlinks the raw observation and keeps the satellite light at the cost of substantial bandwidth, while fully onboard execution avoids the large downlink but loads the satellite with the full computation and energy cost. Binary offloading selects one of the two extremes according to the available resources and exposes the trade-off between onboard load and transmission cost that was studied early in collaborative inference \cite{kang2017neurosurgeon}. Coarse-grained offloading has low control complexity, but it cannot exploit the large layer-to-layer variation in computing cost and intermediate-feature size.

Fine-grained partitioning instead places the split inside the network. The satellite runs the early layers, the ground runs the remaining layers, and only the intermediate features crossing the split point are transmitted. A shallow split keeps the onboard load small but sends larger features, while a deep split reduces the payload at the cost of more onboard computation and energy. The split point therefore couples onboard computation, downlink volume, and ground computation, and choosing it is a joint resource allocation problem rather than a purely architectural decision. Because the split decision is discrete whereas bandwidth, power, and computing allocations are continuous, practical schemes typically optimize them in separate but coupled steps \cite{liang2024jointdnn}.

Collaborative inference is beneficial only when the reduction in onboard computation outweighs the additional feature-transmission and ground-processing cost. The optimal split therefore changes with link quality, available onboard computation, and latency requirements.

The three modes are not mutually exclusive. A satellite may, for example, complete a task onboard but additionally send uncertain regions for ground refinement, or compress intermediate features before transmitting them in a collaborative-inference mode.

\section{Case Study}

\begin{figure*}[htbp]
	\centering
	\begin{tabular}{@{}ccc@{}}
	\includegraphics[width=0.308\textwidth]{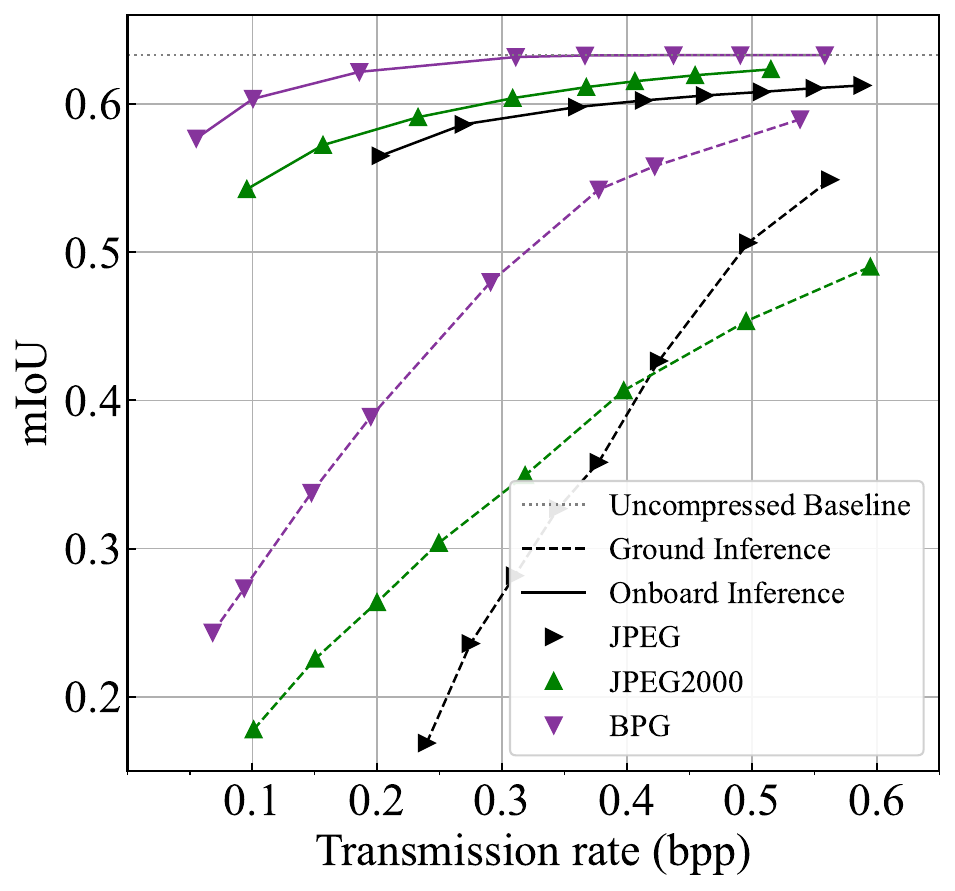} &
	\includegraphics[width=0.32\textwidth]{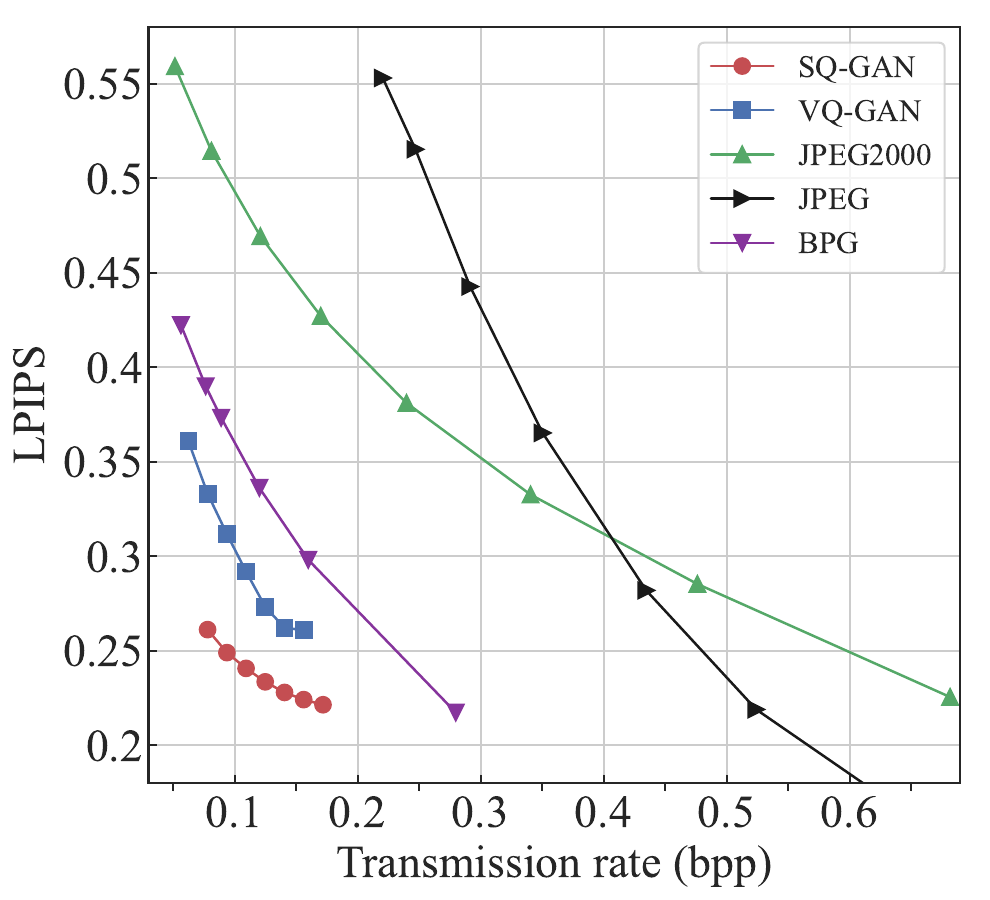} &
	\includegraphics[width=0.32\textwidth]{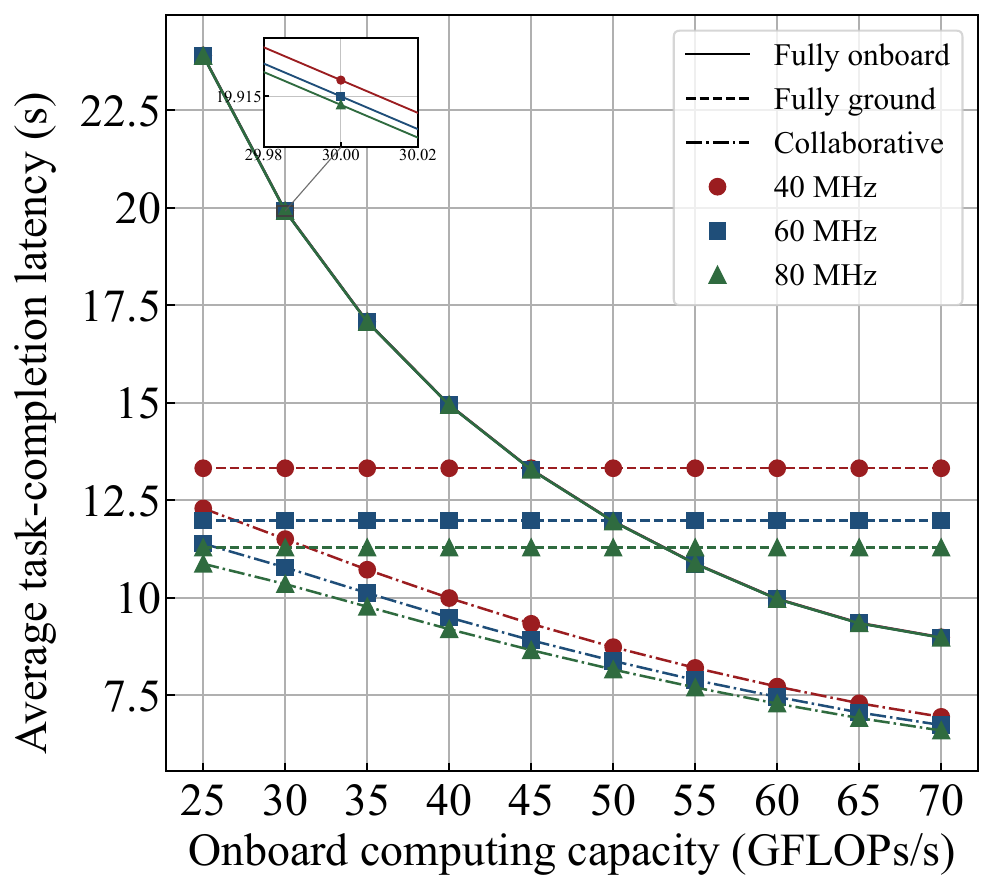} \\
	\footnotesize (a) Onboard task inference. &
	\footnotesize (b) Semantic representation delivery. &
	\footnotesize(c) Space-ground collaborative inference.
	\end{tabular}
	\caption{Performance evaluation of the three operating modes: onboard task inference under downlink constraints, semantic representation delivery under rate constraints, and space-ground collaborative inference under computation constraints. (a) segmentation accuracy versus transmission bitrate; (b) perceptual reconstruction quality versus transmission bitrate; and (c) task-completion latency versus onboard computing capacity under different downlink bandwidths. }
	\label{fig:casestudy}
\end{figure*}

We consider a common land-cover recognition scenario and instantiate the proposed framework under three representative regimes of task requirements and resource constraints. The first two experiments consider constrained downlink transmission but differ in the information required on the ground. Onboard task inference is evaluated in terms of segmentation accuracy when task-level results are sufficient, whereas semantic representation delivery is assessed through reconstruction quality when richer scene information must be retained. The third experiment examines space-ground collaborative inference under varying onboard computing capacities and downlink bandwidths, focusing on task-completion latency. All experiments use the WHDLD dataset, which contains 4{,}940 images of $256\times256$ pixels labeled with six land-cover categories, i.e., building, road, pavement, vegetation, bare soil, and water. The images are split into training, validation, and test subsets at 8:1:1.

For onboard task inference, we use a ResNet-50 backbone with a LoG-CAN classifier head to produce pixel-level label maps and evaluate them by mean intersection over union (mIoU) \cite{he2016deep, ma2023LogCan}. For semantic representation delivery, each image is encoded into a $32\times32$ grid of 1{,}024 discrete tokens from a 1{,}024-entry codebook, and the reconstruction quality is measured by learned perceptual image patch similarity (LPIPS) \cite{esser2021taming, pezone2025sq}. For space-ground split inference, 12 low Earth orbit satellites under a $10^{\circ}$ minimum elevation-angle constraint access a single ground station with 0.8 TFLOPs/s computing capability, and the split point is jointly optimized with communication and computing resources \cite{liang2024jointdnn}. The available bandwidth is orthogonally allocated among satellites, and each satellite serves 24 inference tasks on average, subject to a maximum transmit power of 20 W and an energy budget of 200 J per scheduling period. The satellite-ground link adopts a 12 GHz Ku-band carrier at an altitude of 600 km with atmospheric attenuation.

Fig.~\ref{fig:casestudy} evaluates each mode using a metric matched to its role: segmentation accuracy, reconstruction quality, and task-completion latency; hence, each one reports how the corresponding mode behaves when its own bottleneck is the binding constraint.
\subsection{Onboard Task Inference under Downlink Constraints}
The first experiment compares two ways of using the available downlink rate: transmitting a compressed remote sensing image for ground-side segmentation, or performing segmentation onboard and transmitting only the resulting semantic map. As shown in Fig.~\ref{fig:casestudy}(a), the difference becomes pronounced in the low-rate region.
When the original image is compressed aggressively, segmentation accuracy degrades rapidly because compression artifacts remove visual details that are important to the task. In contrast, once segmentation is completed onboard, the downlink carries the task result rather than the full visual content, allowing many more bits to be used for transmitting task-relevant information.
At a transmission rate of 0.1 bits per pixel (bpp), delivering the onboard segmentation map achieves an mIoU of 0.60 with BPG and 0.54 with JPEG2000, compared with only 0.28 and 0.18, respectively, when the compressed source image is sent to the ground for decision. As the rate increases, the semantic map delivery scheme approaches the uncompressed-image upper bound of 0.633, while raw-image transmission remains more sensitive to compression. Onboard task inference achieves higher segmentation accuracy than image transmission followed by ground-side inference under the evaluated bit budgets. By transmitting compact task-level outputs, it reduces the amount of visual information that must be delivered to the ground. This approach is suitable when the required task is known in advance and the onboard model can produce sufficiently reliable results. 
\subsection{Semantic Representation Delivery under Rate Constraints}
The second experiment considers a different requirement, where the ground needs more information than a final classification or segmentation result, and hence transmitting only a task-level output is insufficient. The comparison therefore focuses on how efficiently different schemes preserve scene information as the transmission rate decreases.
Fig.~\ref{fig:casestudy}(b) shows that token-based representations are particularly effective in the low-rate regime. The unified token scheme, based on VQ-GAN, already outperforms conventional image codecs because it replaces dense pixel transmission with a compact discrete representation. Semantic-aware token selection further improves this tradeoff by allocating the bit budget preferentially to visually and semantically important regions. For example, SQ-GAN achieves an LPIPS of 0.261 at 0.078 bpp. Reaching a similar reconstruction quality requires approximately 0.15 bpp with VQ-GAN, 0.21 bpp with BPG, and 0.55 bpp with JPEG2000. At very low rates, conventional codecs experience substantial quality degradation, with LPIPS exceeding 0.42 below 0.06 bpp, whereas semantic-token transmission maintains useful reconstruction quality at much lower rates.

Semantic-aware token transmission achieves lower perceptual reconstruction error than the evaluated alternatives at low bitrates. Unlike onboard task inference, which delivers task-specific outputs, this mode retains a richer representation of the observed scene for ground-side reconstruction and further processing. Its effectiveness depends on whether the selected tokens preserve information relevant to subsequent applications. 
\subsection{Space-Ground Collaborative Inference under Computation Constraints}
The third experiment shifts the bottleneck from communication to onboard computation. We compare three execution strategies: fully onboard inference, fully ground inference, and model partitioning between satellite and ground.
Their latency trends in Fig.~\ref{fig:casestudy}(c) are fundamentally different. Fully onboard execution avoids feature transmission but is strongly limited by satellite computing capability. Fully ground execution removes most onboard computation, but its latency is dominated by the space-ground link and therefore benefits mainly from additional bandwidth. Collaborative inference operates between these two extremes by selecting a split point that balances onboard computation against the size of the transmitted intermediate features.

Under an 80 MHz bandwidth, increasing satellite computing capability from 25 to 70 GFLOPs/s reduces fully onboard latency from 23.7 s to 8.8 s. Fully ground execution changes much less with onboard capability and remains primarily constrained by transmission, with latency around 13.2 s, 12.2 s, and 11.4 s under 40, 60, and 80 MHz, respectively. By jointly adapting the model split and resource allocation, the collaborative scheme reaches 6.7 s in the 80 MHz case and reduces latency by up to 54.5\% and 47.9\% relative to fully onboard and fully ground execution, respectively.

The comparison confirms that the preferred execution point varies with both onboard computing capability and space-ground transmission conditions. Collaborative inference is beneficial only when the reduction in onboard processing time offsets the additional cost of feature transmission and ground-side completion, making the split point inherently resource-dependent.

\section{Challenges and Future Research Directions}

The three operating modes provide different ways to distribute remote sensing processing between satellites and ground infrastructure. Their practical deployment, however, depends on the reliability of onboard semantic information, the ability to adapt processing decisions to changing resources, and coordination across distributed satellite and ground nodes. 

\subsection{Task-Adaptive and Trustworthy Semantic Extraction}
Semantic extraction faces substantial uncertainty when applied to geographic regions, seasons, sensors, or object categories that differ from the training data. Such uncertainty is particularly consequential in task result delivery, where the original observation may not be available for ground-side verification. A confidently misclassified region or missed object can therefore lead to information loss that cannot be corrected through subsequent processing. 

An important research problem is how to quantify the reliability of extracted semantics and incorporate this information into transmission decisions. Rather than relying solely on average segmentation or detection accuracy, onboard models should estimate prediction uncertainty at the region or object level. These estimates could guide selective transmission of uncertain image patches, additional protection of critical information, or ground-side refinement. However, reliable uncertainty estimation remains difficult under distribution shifts, especially when onboard computing and labeled data are limited. 

Semantic extraction must also accommodate changing mission requirements. Models trained for a fixed set of land-cover classes may be unsuitable for unexpected objects or newly specified observation targets. Open-vocabulary models distilled from ground-based foundation models \cite{sang2026onboard} offer possible solutions, yet adapting them to constrained onboard hardware while preserving task accuracy remains challenging. 

\subsection{Dynamic Communication-Computing Orchestration under Intermittent Connectivity}
The preferred processing mode depends on the available communication and computing resources, which may vary substantially during a satellite pass. Downlink capacity changes with satellite-ground geometry, the remaining contact window continuously decreases, and onboard computing and energy resources are shared among competing tasks. Consequently, a processing mode selected at task arrival may no longer be suitable when the data are ready for transmission. 

A central challenge is to jointly optimize processing mode, model partitioning, and resource allocation under time-varying connectivity. Such decisions involve discrete choices, including the operating mode and model split point, together with continuous variables such as bandwidth, transmission power, and computing allocation \cite{liang2024jointdnn}. Their coupling becomes more difficult when the system must account for future contact opportunities and task deadlines. 

Another issue is incomplete transmission. A contact window may close before all semantic tokens or intermediate features have been delivered. This motivates representations and transmission policies that preserve useful task information under partial reception, together with scheduling strategies that account for the remaining contact time. The resulting optimization should consider task completion probability and energy consumption, in addition to conventional transmission rate and processing latency. 

\subsection{Constellation-Scale Closed-Loop Space-Ground Intelligence}
Extending collaborative inference from a single satellite-ground pair to a constellation introduces new coordination requirements. Remote sensing observations, computing resources, and model components may be distributed across multiple satellites and ground stations. A task may involve observations from several platforms, while different satellites may operate with independently updated semantic encoders or inference models. 

One challenge is to maintain consistency among distributed models and semantic representations. Differences in model versions, codebooks, or feature representations can affect whether information generated by one satellite can be processed by another node. Model synchronization and update scheduling must therefore account for intermittent inter-satellite and space-ground links, limited onboard storage, and the communication cost of distributing model parameters. 

Ground-side feedback provides another opportunity for improving onboard intelligence. Corrections to segmentation results, missed detections, and newly annotated observations can support model adaptation and updates to semantic importance estimation. However, determining which information should be returned to orbit, which satellites require an update, and when updates should be delivered remains an open resource-allocation problem. 

\section{Conclusion}
This article presented a semantic-aware space-ground collaborative remote sensing framework that supports different processing modes according to task requirements and communication-computing constraints. The three modes differ in the amount of onboard inference, the representation transmitted over the space-ground link, and the processing required on the ground. The case study illustrates their respective advantages in task-result delivery, low-rate semantic representation transmission, and latency-aware collaborative inference. These insights emphasize both the promise and the challenges of in-orbit intelligence, providing a strategic roadmap for task- and resource-aware space-ground remote sensing.


\balance
\bibliographystyle{IEEEtran}
\bibliography{references}

\end{document}